\documentstyle[12pt,epsf]{article}
\global\arraycolsep=2pt
\def\bc{\begin{center}}
\def\ec{\end{center}}
\def\be{\begin{equation}}
\def\ee{\end{equation}}
\def\bea{\begin{eqnarray}}
\def\eea{\end{eqnarray}}

\def\GeV{\mathord{\rm \;GeV}}

\def\sinh{\mathop{\rm sinh}\nolimits}

\def\rsim{\mathrel{\raise2pt\hbox to 8pt{\raise -5pt\hbox{$\sim$}\hss{$>$}}}}
\def\lsim{\mathrel{\raise2pt\hbox to 8pt{\raise -5pt\hbox{$\sim$}\hss{$<$}}}}

\def\vev#1{\langle #1\rangle}
\def\sbar{\bar s}
\def\CONT{{\em cont}}
\def\LATT{{\em lat}}
\def\UN{{\rm UN}}
\def\SM{{\rm SM}}
\def\GI{{\rm GI}}
\def\DREZp{\hbox{${\rm DR\overline{EZ}}'$}}
\def\NDR{{\rm NDR}}
\def\MSBAR{{\hbox{$\overline{\rm MS}$}}}
\def\MSbar{{\overline{\rm MS}}}
\def\opoldc#1#2#3{\,[#1\times#2]_{#3}}

    \def\CO{{\cal O}} 
\def\CQ{{\cal Q}}

\def\NPB#1{{\it Nucl. Phys.} {\bf B#1}}
\def\NPBPS#1{{\it Nucl. Phys.} {\bf B} ({\it Proc. Suppl.}) {\bf #1}}
\def\PRL#1{{\it Phys. Rev. Lett.} {\bf #1}}

\def\PRD#1{{\it Phys. Rev.} {\bf D#1}}

\begin{document}

\begin{titlepage}

\begin{flushright}
\vspace{-0.3truein}
LAUR-97-2134\\
UW/PT-97-16
\end{flushright}


\begin{center}
{\LARGE\bf Staggered fermion matrix elements \\
using smeared operators}

\end{center}

\vspace{0.2cm}

\begin{center}
 {
  \begin{tabular}[t]{c}
	Greg Kilcup \footnotemark\\
	\em Physics Department, The Ohio State University, 
	Columbus, OH 43210 \\[1em]
        Rajan Gupta \footnotemark\\
        \em Los Alamos National Laboratory, Mail Stop B-285,
        Los Alamos, NM 87545\\[1em]
        Stephen R. Sharpe \footnotemark\\
        \em Physics Department, Box 351560, University of Washington,
        Seattle, WA 98195-1560 \\
  \end{tabular}}
\end{center}

\vspace{0.2cm}

\begin{small}
\centerline{ABSTRACT}
\medskip

  We investigate the use of two kinds of staggered fermion operators, 
  smeared and unsmeared. The smeared operators extend over a $4^4$ hypercube,
  and tend to have smaller perturbative corrections than the corresponding
  unsmeared operators. We use these operators to calculate kaon weak matrix 
  elements on quenched ensembles at $\beta=6.0$, $6.2$ and $6.4$.
  Extrapolating to the continuum limit, we find $B_K(NDR, 2 \GeV)=
  0.62\pm0.02({\rm stat})\pm0.02({\rm syst})$. The systematic error
  is dominated by the uncertainty in the matching between lattice and 
  continuum operators due to the truncation of perturbation theory
  at one-loop. We do not include any estimate of the errors due to quenching
  or to the use of degenerate $s$ and $d$ quarks.
  For the $\Delta I = {3/2}$ electromagnetic penguin 
  operators we find  $B_7^{(3/2)} = 0.62\pm0.03\pm0.06$ and 
  $B_8^{(3/2)} = 0.77\pm0.04\pm0.04$. 
  We also use the ratio of unsmeared to smeared operators to make a partially
  non-perturbative estimate of the renormalization of the quark mass
  for staggered fermions. We find that tadpole improved perturbation
  theory works well if the coupling is chosen to be $\alpha_\MSbar(q^*=1/a)$.
  
\end{small}

\bigskip
\centerline{(submitted to Physical Review D)}

\footnotetext[1]{Email: kilcup@physics.ohio-state.edu}
\footnotetext{Email:    rajan@qcd.lanl.gov}
\footnotetext{Email:    sharpe@phys.washington.edu}
\vfill
\mbox{July, 1997}
\end{titlepage}

\section{Introduction}
\label{sec:intro}

An important goal of lattice QCD is to provide reliable calculations
of electroweak matrix elements.
The major sources of error in present calculations are the use of
finite lattice spacing, the use of one-loop perturbation theory
to match continuum and lattice operators, and the use of the quenched
approximation \cite{sslat97}. 
In this paper we address the first two errors for
calculations using staggered fermions. 
In particular, we test the efficacy of ``smeared'' operators~\cite{PS}.
These extend over a $4^4$ hypercube, and thus are larger than
the usual (``unsmeared'') operators which are confined to a $2^4$ hypercube.
Nevertheless, in many cases they are closer to the continuum operators
in the sense that the one-loop matching coefficients are closer to unity.

We apply these operators to the study of three quantities. 
The first is the kaon B-parameter, $B_K$,
which we study using both smeared and unsmeared operators. 
The initial motivation for introducing smeared operators was the discovery
of large discretization errors in the results for unsmeared 
operators~\cite{sslat90,sslat91}.
At the time, it was unclear whether the discretization errors were
proportional to the lattice spacing $a$ or to $a^2$ 
(up to logarithmic corrections).
The smeared operators were designed to reduce possible $O(a)$ errors---they
match onto continuum operators with no errors of $O(a)$ at tree level.
It was subsequently realized that the $O(a)$ parts of the lattice operators 
do not contribute to $B_K$, and that the errors are automatically of $O(a^2)$
for both the unsmeared and smeared operators \cite{sslat93,luo,jlqcd96}.
This theoretical argument has since been tested 
numerically~\cite{jlqcd96,kilcup96}.
Thus the interest in using the smeared operators is that they provide
an estimate of the error in the matching of continuum and lattice operators.
The results from different operators
should differ at finite $a$, but agree upon extrapolation to $a=0$,
up to higher order perturbative corrections.

Preliminary results from this study of $B_K$
were presented in Ref.~\cite{sslat93},
and one of our purposes here is to present final results.
Although, by present standards, these come from a small statistical 
sample, the errors are nevertheless small enough to assess 
the impact of smeared operators. 
We have also improved our estimates of the error due to the truncation of
the perturbative matching factors, using the method introduced in 
Ref.~\cite{bkwnew}.

Our second application is the calculation of $B_7^{3/2}$ and $B_8^{3/2}$.  
These B-parameters, and in particular $B_8^{3/2}$, determine the size of the
electromagnetic penguin contribution to $\epsilon'$.  
In contrast to $B_K$,
the use of staggered fermions for these quantities offers no clear
advantage over Wilson fermions.
However, since the systematic errors in the results with the two types of
fermion are different, an important check of the reliability of the
lattice calculations is to show that the two formulations give 
consistent results in the continuum limit.  To this end we compare the
staggered data with the recent results obtained using Wilson fermions
at $\beta=6/g^2=6$ in the quenched approximation~\cite{bkwnew}.

The calculation of $B_7^{3/2}$ and $B_8^{3/2}$ demonstrates
the importance of using several discretizations of continuum operators.
It turns out that one cannot use the unsmeared
operators because the one-loop correction to the matching coefficients
approaches 100\% \cite{SP}.  On the other hand, the one-loop
corrections are much smaller for the smeared operators ($\sim25\%$), 
and we can use them to calculate $B_7^{3/2}$ and $B_8^{3/2}$.
The only way in which one could use the unsmeared operators would be to
develop a non-perturbative method of determining the
matching coefficients (which is possible in principle using external
quark states~\cite{nonpertzs}).

Our final application concerns the calculation of the matching
relation between the lattice and continuum regularization schemes,
particularly in cases where the reliability of perturbative estimates
is questionable.  The ratio of such matching factors for two different
discretizations of an operator can be estimated non-perturbatively by
taking ratios of appropriate matrix elements.  The non-perturbative
results so obtained can be used to test the reliability of the
one-loop perturbative estimates.  In particular, one can use the results
to fix the scale $q^*$ at which to evaluate the
coupling constant entering into the perturbative expressions.
For the pseudoscalar density we find that
$q^* \approx 1/a$. Our
conclusions are, however, preliminary, since we do not have results at
enough values of lattice spacing to check the extrapolations we use
to remove discretization errors.

We use this method to assess the reliability of the matching factor,
$Z_m$. $Z_m$ relates the bare lattice mass
to the continuum mass in, say, the $\MSBAR$ scheme, and is a crucial
ingredient in the calculation of continuum light quark masses from the
lattice.  Recent work has suggested that continuum quark masses are smaller
than previously thought, but this is based
on trusting one-loop perturbation theory for $Z_m$ \cite{GB}.
For staggered fermions,
the one-loop contribution to $Z_m$ is large, roughly a $60\%$
correction at $\beta=6$, even after tadpole improvement \cite{PS}.
This casts doubt on the reliability of the perturbative $Z_m$, and
therefore also on the extracted value of quark masses.  We find,
however, that our partly non-perturbative estimate suggests that the
one-loop evaluation is close to the correct answer.  

The organization of this paper is as follows. In the following section
we describe the method we use to match continuum operators 
to lattice operators composed of staggered fermions.
In sec.~\ref{sec:details} we give a short
description of the numerical methods and data sample.
The three subsequent sections contain
our analysis and results for $B_K$, $B_{7,8}^{3/2}$, and the
non-perturbative ratios of matching factors, respectively. 
We close with some conclusions.

\section{Theoretical Review}
\label{sec:theory}

In this section we explain our method of calculating $B$-parameters using
staggered fermions. This requires combining a variety of results
already in the literature, and we focus here only on the essential
details. For a more extensive description of the method
see Ref.~\cite{sstasi}.

The continuum operators of interest are
\begin{eqnarray}
\CQ_K &=& \sbar_a\gamma_\mu^L d_a\, \sbar_b\gamma_\mu^L d_b  
\,,\label{eq:QK}\\
  \CQ_7^{3/2}
  &=& \bar s_a \gamma_\mu^L d_a \  \big[ 
    \bar u_b \gamma_\mu^R u_b - 
    \bar d_b \gamma_\mu^R d_b \  \big] + 
    \bar s_a \gamma_\mu^L u_a\, \bar u_b \gamma_\mu R d_b
\,,\label{eq:Q7}\\
%
%
  \CQ_8^{3/2}
&=& \bar s_a \gamma_\mu^L d_b \  \big[
    \bar u_b \gamma_\mu^R u_a - 
    \bar d_b \gamma_\mu^R d_a \  \big] + 
    \bar s_a \gamma_\mu^L u_b \, \bar u_b \gamma_\mu R d_a \,.  
\label{eq:Q8}
\end{eqnarray}
where $a,b$ are color indices and 
$\gamma_\mu^{R,L}=\gamma_\mu(1\pm\gamma_5)$.  All operators
are in Euclidean space, and we use hermitian Gamma matrices.  
The superscripts on $\CQ_{7,8}$ indicate that these are the $I=3/2$
parts of the operators $\CQ_{7,8}$, 
i.e. the $I=1/2$ component has been removed. 
We make this restriction because the calculation of the
matrix elements of the $I=3/2$ parts is much simpler.
In the limit of exact flavor $SU(3)$, which is the limit we work in here,
the $I=3/2$ parts give rise only to ``eight'' diagrams, i.e. those in
which the quark fields in the operator are contracted with fields in the
external mesons. These are the same type of diagrams which contribute to the
matrix element of $\CQ_K$.
The $I=1/2$ parts, by contrast, give rise also to ``penguin'' or ``eye''
diagrams, which are much more difficult to calculate.
The restriction to the $I=3/2$ parts of $\CQ_{7,8}$ does not, however,
diminish the phenomenological interest in the results,
because $\CQ_{7,8}^{3/2}$ are the only operators which give an
imaginary part to the $K^+\to\pi^+\pi^0$ amplitude.

To form $B$-parameters we also need the matrix elements of the axial
and pseudoscalar densities,
\begin{equation}
A_\mu = \bar s_a \gamma_\mu \gamma_5 d_a \,, \qquad
P = \bar s_a \gamma_5 d_a \,. 
\end{equation}
We can then define
\begin{eqnarray}
B_K &=& { \vev{{K^0}|\CQ_K| \bar{K}^{0}} \over
(8/3)\vev{{K^0}|A_4 |0}\, \vev{0| A_4 | \bar{K}^{0}} }
\,,\label{eq:BKdef}\\
B_7^{3/2} &=& {\vev{\pi^+|\CQ_7^{3/2}| K^+} \over
(2/3)\vev{{K^0}|P|0}\, \vev{0| P| \bar{K}^{0}} -
\vev{{K^0}|A_4 |0}\, \vev{0| A_4 | \bar{K}^{0}} }
\,,\label{eq:B7def}\\
B_8^{3/2} &=& {\vev{\pi^+|\CQ_8^{3/2}| K^+} \over
2\vev{{K^0}|P|0}\, \vev{0| P| \bar{K}^{0}} - (1/3)
\vev{{K^0}|A_4 |0}\, \vev{0| A_4 | \bar{K}^{0}} }
\,.\label{eq:B8def} 
\end{eqnarray}
All external particles have been assumed to be at rest.  For brevity,
we have used $SU(3)$ flavor symmetry to rewrite all the denominators
in terms of kaon matrix elements.  Note that, in general, both
numerators and denominators of these ratios depend upon the
renormalization scale $\mu$ and the scheme used to define the operators.  
When quoting physical values we use the NDR scheme,
i.e. $\MSBAR$ renormalization combined with a particular set of rules for
treating $\gamma_5$ away from four dimensions,
and choose the renormalization scale to be $\mu = 2$ GeV.

The extraction of the above matrix elements using 
staggered fermions is complicated by the mixing between the
spin and flavor degrees of freedom.  As explained in
Refs.~\cite{SP,sstasi}, we proceed in two stages.
We first match the continuum matrix elements onto those in an
``enlarged'' continuum theory, 
and then match from that theory onto the lattice.
The enlarged theory differs from QCD by having eight copies of each
physical quark.
The eight copies of the strange quark spinor are collected into two
$4\times4$ matrices $S_{\beta,b}$ and $S'_{\beta,b}$, where $\beta$ is
a spinor index, and $b=1-4$ is a ``staggered-flavor'' index. Similar
fields are constructed for the up and down quarks. 
The correspondence between matrix elements in the continuum and in
the enlarged theory is
\begin{eqnarray}
\vev{0| A_4 | \bar{K}^{0}} &=& \sqrt{1\over N_f}
\vev{0| \bar S (\gamma_4\gamma_5\otimes\xi_5) D| \bar{K}^{0}_{G}} 
\,, \label{eq:correpA}\\
\vev{K^{0}| A_4 | 0} &=& \sqrt{1\over N_f}
\vev{K'^{0}_{G}| \bar S' (\gamma_4\gamma_5\otimes\xi_5) D'| 0} 
\,, \label{eq:correpAp}\\
\vev{0| P| \bar{K}^{0}} &=& \sqrt{1\over N_f}
\vev{0| \bar S (\gamma_5\otimes\xi_5) D| \bar{K}^{0}_{G}} 
\,, \label{eq:correpP}\\
\vev{K^{0}| P | 0} &=& \sqrt{1\over N_f}
\vev{K'^{0}_{G}| \bar S' (\gamma_5\otimes\xi_5) D'| 0} 
\,, \label{eq:correpPp}\\
\vev{{K^0}|\CQ_K| \bar{K}^{0}} &=& {2\over N_f}
\vev{K'^{0}_{G}|\, 
	[(V-A)\times P]_{I} + [(V-A)\times P]_{II}\,|\bar{K}^{0}_{G}}
\,, \label{eq:correpMK}\\
\vev{\pi^+|\CQ_7^{3/2}| K^+} &=& {1\over N_f}
\vev{K'^{0}_{G}|\, 2\,[(P-S)\times P]_{I} + [(V+A)\times P]_{II}\,
	|\bar{K}^{0}_{G}}
\,, \label{eq:correpM7}\\
\vev{\pi^+|\CQ_8^{3/2}| \bar{K}^+} &=& {1\over N_f}
\vev{K'^{0}_{G}|
\, 2[(P-S)\times P]_{II} + [(V+A)\times P]_{I}\,|\bar{K}^{0}_{G}}
\,. \label{eq:correpM8}
\end{eqnarray}
We have used flavor symmetry to
rewrite all matrix elements in terms of those between external kaon states.
The notation for matrix elements in the enlarged theory is that of
Refs.~\cite{PS,SP}, and we give only a brief summary.
The matrices appearing in bilinears are tensor products of
spin and staggered-flavor matrices.
For example, $(\gamma_5\otimes\xi_5)$ indicates a pseudoscalar density
with staggered-flavor matrix $\gamma_5$.
The states $|\bar{K}^{0}_{G}\rangle$ and $|\bar K'^{0}_{G}\rangle$ are
those created from the vacuum by $\bar D (\gamma_5\times\xi_5) S$ 
and $\bar D' (\gamma_5\times\xi_5) S'$, respectively.
The states are normalized, which leads to the factors of
$N_f=4$, the usual multiplicity factor for staggered fermions.
The subscript $G$ indicates that these states
are the pseudo-Goldstone bosons corresponding to the axial $U(1)$
symmetry which is unbroken when one discretizes the enlarged theory
using staggered fermions and takes the chiral limit.
The notation for four-fermion operators is from Ref.~\cite{SP}.
For example, $[(V-A)\times P]_{I}$ represents the one color loop 
contraction of the four-fermion operator with spin structure
$\gamma_\mu\cdot\gamma_\mu - \gamma_\mu\gamma_5\cdot\gamma_5\gamma_\mu$,
and in which both bilinears have staggered flavor $\gamma_5$. 
Finally, the factor of $2$ on the r.h.s. of Eq.~(\ref{eq:correpMK}) arises
from the difference in the number of Wick contractions in QCD and the
enlarged theory due to the use of ``primed'' quarks in the latter theory.

All these equalities hold identically between the quenched versions of the
two theories.
In the presence of internal fermion loops, they hold if,
in the enlarged theory, each loop is multiplied by $1/8$, i.e. 
the fermion determinant is taken to the power $1/8$.

The next step is to relate the operators in the enlarged continuum theory to
those in the corresponding lattice theory. The enlarged theory has been
chosen so that it is the continuum limit of a discretized theory in
which one uses a single staggered species for each $4\times4$
matrix field, e.g. $\chi_{S}$ for $S$ and $\chi_{S'}$ for $S'$.
This means that the relation between operators is simple at tree level.
For example, using the notation of Ref.~\cite{PS},
\begin{equation}
\bar S (\gamma_5\otimes\xi_5) D = 
\bar\chi_S \overline{(\gamma_5\otimes\xi_5)} \chi_D\,
 [1 + O(a) + O(\alpha_s)] \,.
\label{eq:bilmatch}
\end{equation}
Here $\overline{(\gamma_5\otimes\xi_5)}$ is that matrix in the space of
possible positions in a $2^4$ hypercube which corresponds to the
continuum spin-flavor matrix $(\gamma_5\otimes\xi_5)$.
Similar relations hold for the four fermion operators \cite{SP}.


In this paper we use one-loop perturbation theory to match operators
in the enlarged continuum theory to those on the lattice.
In other words, we include
the terms of $O(\alpha_s)$ in equations such as (\ref{eq:bilmatch}).
More precisely, we use the ``horizontal matching'' procedure discussed 
in detail in Ref.~\cite{bkwnew}.  
This consists of two steps.
We first match the lattice and continuum operators at an
intermediate scale $q^*\sim 1/a$ using
\begin{equation}
\CO^{\CONT}_i(q^*) = \CO^{\LATT}_i  + {\alpha_\MSbar(q^*)\over 4\pi} 
\sum_j \left(-\gamma_{ij}^{(0)} \ln(q^* a) + c_{ij}\right)  
\CO^{\LATT}_j + O(\alpha^2) + O(a)\,.
\label{eq:contlatt}
\end{equation}
Here $\CO^\LATT_i$ are bare lattice operators,
$\gamma^{(0)}$ is the one-loop anomalous dimension matrix,
and the $c_{ij}$  are the one-loop matching coefficients.
We give, below, numerical values for the $c_{ij}$ of interest.
The second step is to evolve the result 
from $q^*$ to the final scale $\mu=2\,$GeV using the continuum 
two-loop anomalous dimension matrix.
The two-loop anomalous dimensions for the continuum operators of interest are
collected in Ref. \cite{bkwnew}, and we do not repeat them here.

The point of this procedure is to account for the fact that there
are two scales in the problem: the matching scale $q^*\sim 1/a$,
and the final scale $\mu$. These can differ substantially---indeed the ratio
$q^*/\mu$ can be as large as $10$ in our calculation. 
The coupling constant can thus be quite different at the
two scales, and it is important to make sure that the appropriate
coupling is used at each stage. Furthermore, it becomes
necessary to sum up the leading logarithms of $q^*/\mu$.
Both of these requirements are accomplished by horizontal matching.
The scale in the coupling in the first step should be $q^*$,
and the renormalization group evolution in the second step sums up the
leading logarithms.
Of course, since we truncate perturbation theory, 
there are errors in both steps. 
In the matching relation~(\ref{eq:contlatt}) 
the truncation errors are of $O[\alpha_s(q^*)^2]$,
while the errors in the continuum evolution are proportional to both
$O[\alpha_s(q^*)^2]$ and $O[\alpha_s(\mu)^2]$.
We note, however, that the errors from the continuum
evolution are the same for any choice of lattice 
discretization of the continuum operator.
This point will be important in our discussion of results for $B_K$
using different lattice operators.\footnote{%
For further discussion of horizontal matching,
and in particular of its relation to the exact matching formula of 
Ji~\cite{Ji}, see Ref.~\cite{bkwnew}.}

To use horizontal matching we need to choose the
intermediate matching scale $q^*$.
If one worked to all orders in perturbation theory, the final result
would be independent of $q^*$.  
This is not true if one truncates perturbation theory:
different choices lead to results differing by $O[\alpha(q^*)^2]$.
One should choose $q^*$ to be, roughly speaking, the average momentum
flowing through the quark-gluon vertices in the matching 
calculation~\cite{lepage}.
It is difficult, however, to make this into a precise prescription
for matching calculations involving operators with non-zero anomalous
dimensions. The only method we know of was proposed in Ref.~\cite{bkwnew},
and involves applying the BLM prescription of Ref.~\cite{BLM} to the
calculation of the two-loop lattice anomalous dimension matrix.
This calculation has not been done, and so we have chosen to use a
range of values for $q^*$.
Since the lattice and continuum operators are constructed to have the
same matrix elements at long distances, the dominant contributions to
the matching calculation are from momenta near the lattice cut-off
(as long as the continuum renormalization point is also of this size).
Thus we take $q^*=K/a$ with $K$ a constant.
Since we use tadpole improved operators we expect that $K\approx 1$
rather than $K\approx \pi$, as explained in Ref.~\cite{lepage}.
So we have chosen $q^*=1/a$ for our central values, and used $q^*=\pi/a$
in order to estimate the uncertainty due to the truncation of
perturbation theory.\footnote{%
Reference~\cite{lepage} advocates the use of a 
different coupling constant, $\alpha_V$, rather than $\alpha_\MSbar$. 
$\alpha_V$ is defined in terms of the quark-antiquark potential.
We prefer to use $\alpha_\MSbar$ because this is the scheme used in continuum
calculations of coefficient functions, and the matching formulae are simpler if
one uses the same coupling in the continuum and on the lattice.
The two couplings are quite similar---our choice of $q^*=1/a$ in the $\MSBAR$
scheme corresponds to using $\alpha_V(1.6/a)$. 
This is indeed a typical value for $q^*$ for tadpole improved quantities.}
This is certainly a crude estimate, but it is the best that we can do.

In our preliminary analysis of $B_K$ we did not use horizontal matching,
but rather used the one-loop matching relation Eq.~(\ref{eq:contlatt})
connect directly from the lattice to the final scale $\mu$ \cite{sslat93}.
To estimate the truncation error we took the difference between the results
of using $\alpha_s(\mu)$ and $\alpha_s(q^*)$ in the matching equation.
Our present methods both of matching and of estimating the truncation error
are more reliable, although, as we will see below, 
the final answer is little changed.
 
The $\MSBAR$ coupling constant appearing in Eq.~(\ref{eq:contlatt}) 
is determined using the method of Ref. \cite{lepage}, 
which incorporates tadpole improvement.
We first solve
\begin{equation}
- \log\Box
= {4\pi\over3} \alpha_V(3.41/a) \left[1 - 1.185 \alpha_V(3.41/a)\right]
\label{eq:alpha1}
\end{equation}
for $\alpha_V$, where $\Box$ is the expectation value of the plaquette
normalized to be unity in the continuum limit.
We then convert to the $\MSBAR$ scheme using
\begin{equation}
\alpha_\MSbar(3.41 / a) = 
\alpha_V(e^{5/6}\; 3.41 / a) (1 + 2\alpha_V / \pi) \,.
\label{eq:alpha2}
\end{equation}
This coupling is then run to other scales using the two-loop 
$\beta$-function.

The lattice operators we use are of two types: the standard operators,
which are contained in $2^4$ hypercubes (and which we refer to as
``unsmeared'' operators); and the ``smeared'' operators introduced in
Ref.~\cite{PS}, which live on $4^4$ hypercubes. 
The smeared operators have the same form as the unsmeared, except for the
replacement
\begin{equation}
  \chi(y+A) \to \frac14 \sum_{\mu=1}^{4} \chi(y + A +2\hat\mu[1-2A_\mu] )
\,,\label{eq:defsmear}
\end{equation}
where $y$ is the position of the origin of the $2^4$ hypercube,
and $A$ is the position within the hypercube.
This definition makes the average position of the quark field lie at the center
of the hypercube, 
which is why there are no $O(a)$ corrections when matching to the
continuum operator at tree level.
Both types of operator are made gauge invariant by fixing to lattice Landau gauge.
Both are also tadpole improved, using the
mean link determined from the plaquette, $u_0^4 =\Box$.

Although the discretization errors in the matching equation are, in
general, of $O(a)$, it turns out that for the matrix elements of
interest the corrections are actually of $O(a^2)$.  This is because
the $O(a)$ parts have the wrong staggered-flavor
\cite{sslat93,ssinprep}, and by using staggered flavor $\xi_5$ for the
incoming and outgoing states we project against them up to errors of $O(a^2)$. 
For the same reason, we need
only include in the matching equations those operators which have the
correct staggered-flavor.  This is fortunate, since typically many operators of
the wrong flavor are needed to match the continuum operators.

The matching matrices $c_{ij}$ for all continuum operators of
interest, and for both smeared and unsmeared lattice operators, can be
determined from the results of Ref.~\cite{SP}.  When doing this one
must be careful to account for the following three issues.  First, the
numerical results for the $c_{ij}$ are quoted in Ref.~\cite{SP}
for continuum operators
defined in the \DREZp, rather than the \NDR, scheme.  Thus one must
convert between these schemes, and a recipe for doing so is given in
\cite{SP}.  Second, tadpole improvement in \cite{SP} is based on the
average trace of the link in Landau gauge, rather than on the average
plaquette which we use here.  This causes a small change in the
coefficients
\begin{equation}
c_{ij} = c_{ij}(\hbox{\rm Ref.~\cite{SP}}) + N_B (4/3)(\pi^2 - 9.17479) 
\delta_{ij} 
\,,
\end{equation}
where $N_B=1$ for bilinears and $N_B=2$ for four fermion
operators. Third, the coefficient of the logarithm in the matching
equation (\ref{eq:contlatt}) has been changed from $q^* a/\pi$ to $q^* a$.
This shifts the $c_{ij}$ of Ref.~\cite{SP} by $\gamma_{ij}^{(0)} \ln\pi$.
This change allows one to see the size of the matching corrections
for our standard value $q^*a = 1$ directly from the size of the $c_{ij}$.

As the procedure for calculating the matching coefficients is rather
involved, we collect here the relevant parts of the final results.
For the bilinears, the matching involves no mixing, only multiplicative
renormalization. The results are
\begin{eqnarray}
c_{A}^{\UN} &=& \ \ 0.9264\,, \quad\qquad c_A^\SM = -1.0120 \,,
\label{eq:Aconnection}\\
c_{P}^{\UN} &=& -39.1414\,, \qquad c_P^\SM = -8.8882 \,.
\label{eq:Pconnection}
\end{eqnarray}
Here $c_A$ is a shorthand for the diagonal coefficient for the axial density,
etc., and the superscript refers to 
whether or not the quark fields are smeared.
These coefficients are multiplied by  $\alpha_\MSbar(q^*)/4\pi$,
which, in our calculation, varies in the range $0.01-0.015$. 
Thus these corrections are numerically small for the axial currents,
($\sim 1\%$), and of moderate size
for the smeared pseudoscalar density ($\sim 10\%$).
For the unsmeared pseudoscalar density, however, they are large
($\sim 50\%$), suggesting that two-loop terms will be important.
This is an example of
the advantage of using smeared operators.

\def\hmi{\hphantom{-}}

For the four-fermion operators of interest here, we can decompose the 
matrices of matching coefficients into two parts.
The first is a square matrix with indices running over operators
\begin{equation}
i,j = \left( \opoldc {(V-A)}PI, \opoldc {(V-A)}P{II}, 
	\opoldc {(V+A)}PI, \opoldc {(V+A)}P{II}\right)
 \,,
\end{equation}
for which we find 
\begin{eqnarray}
c_{ij}^\UN &=&
\left(\begin{array}{llll}
    -4.3079  	& \hmi 0.4611  	& \hmi 0.2638  	&   -0.7913\\
    -1.0002  	& \hmi 0.0761	& \hmi 0      	&   -2.1102\\
   \hmi 0.2638 	&   -0.7913 	&   -8.9740  	&   -1.5387\\
   \hmi 0      	&   -2.1102  	&   -4.9998  	&  \hmi 1.4093
\end{array}\right) 
\label{eq:AVUN}\,, \\
c_{ij}^\SM &=& 
\left(\begin{array}{llll}
    -2.1376  	&   -5.4068  & \hmi 0.0850  	&   -0.2551  \\
    -5.8701  	&   -0.7474  & \hmi 0  		&   -0.6802  \\
 \hmi 0.0850  	&   -0.2551  & \hmi 4.5595  	&   -2.5366  \\
 \hmi 0       	&   -0.6802  &   -0.1299  	&   -2.6608  
\end{array}\right) 
\label{eq:AVSM}\,. 
\end{eqnarray}
These coefficients are needed for all three $B$-parameters, and the
corrections are of moderate size for both smeared and unsmeared operators.
The second matrix we need is rectangular, having indices\footnote{%
Note that, at one-loop, tensor operators do not appear.} 
\begin{eqnarray}
i &=& \left(\opoldc{(P-S)}PI, \opoldc{(P-S)}P{II} \right)\,, \\
j &=& \left(\opoldc{(P-S)}PI, \opoldc{(P-S)}P{II}, 
	      \opoldc{(P+S)}PI, \opoldc{(P+S)}P{II} \right) \,.
\label{eq:PPindices}
\end{eqnarray}
For this we find 
\begin{eqnarray}
c_{ij}^\UN &=&
\left(\begin{array}{llll}
\hmi 0.7562  	&  -14.7309  & \hmi 5.3559  &  -16.0674\\
    -3       	&  -34.4364  & \hmi 0       &  -42.8464
      \end{array}\right) \,,
\label{eq:CijPSUN} \\
c_{ij}^\SM &=&
\left(\begin{array}{llll}
    -2.8862  &   -3.1611  &\hmi 1.6759  &   -5.0277\\
    -3       &   -3.3692  &\hmi 0       &  -13.4072
      \end{array}\right) \,.
\label{eq:CijPS}
\end{eqnarray}
These results are needed for $B_7^{3/2}$ and $B_8^{3/2}$, and it is here
that the use of smeared operators is most important.
In particular, the numbers in the second row in Eq.~(\ref{eq:CijPS}) 
are much smaller than in the corresponding row in Eq.~(\ref{eq:CijPSUN}).
Much of this reduction can be traced back to the similarly large difference 
in the coefficients $c_P$ shown in Eq.~(\ref{eq:Pconnection}).  
For the range of couplings we study, it turns out that the
one-loop matching correction for the operators appearing
in the numerator of $B_7^{3/2}$ and $B_8^{3/2}$ actually exceeds 100\%
if one uses the unsmeared operators with $q^* \approx 1/a$.
Our results for these quantities are, therefore, 
obtained exclusively with smeared operators.

\section{Numerical details}
\label{sec:details}

Our results are based on ensembles of lattices at three different lattice spacings.
A summary of the important parameters is given in Table~\ref{tab:numdetails}.
The lattices at $\beta=6$ are part of a sample previously used to study the
hadron spectrum, and a complete description of how we generated the
lattices and calculated quark propagators can be found in Ref. \cite{oldhm}.
The lattices at $\beta=6.2$ and $6.4$ have been discussed previously
in Refs.~\cite{sslat90,gklat93}. They were generated using overrelaxed and
Metropolis sweeps in a 4:1 ratio, and separated by 1000 sweeps.
They were divided roughly equally into two independent streams at $\beta=6.2$, 
and three at $\beta=6.4$. Quark propagators were calculated using the
conjugate gradient algorithm.

\begin{table}   
\renewcommand{\arraystretch}{1.2}
\begin{center}
\begin{tabular}{lccc}
\hline
$\beta$			& $6.0$ 	& $6.2$		& $6.4$ \\
\hline
Lattice Size		& $24^3\times40$& $32^3\times48$&$32^3\times48$\\
Number of lattices	& 13		&  23		& 24 \\
Samples per lattice	& 1		& 2		& 2 \\
Quark masses		& $0.01$, $0.02$, $0.03$ 
					& $0.005$, $0.01$, $0.015$
						& $0.005$, $0.01$, $0.015$ \\
$1/a$ in GeV from $m_\rho$
			& 1.9		& 2.6		& 3.45	\\
Minimum $M_\pi$ in GeV  & 0.454		& 0.376		& 0.414	\\
Average plaquette	& 0.5937	& 0.6137	& 0.6306 \\
$\alpha_\MSbar(\pi/a)$	& 0.134		& 0.125		& 0.117	\\
$\alpha_\MSbar(1/a)$	& 0.192		& 0.173		& 0.158 \\
$\alpha_\MSbar(2 \GeV)$	& 0.188		& 0.190		& 0.191 \\
$1/a$ for $\alpha_\MSbar(2 \GeV)=0.190$	
			& 1.94		& 2.6		& 3.4	\\
\hline
\end{tabular}
\caption{Parameters of numerical simulations used in this analysis.}
\label{tab:numdetails}
\end{center}
\end{table}

We consider here only pions composed of two degenerate quarks.  For
the quark masses that we use, the masses of the lattice pions bracket
that of the physical kaon at each $\beta$.  To illustrate this we
include in the table the mass of the lightest pion, converted to
physical units using $1/a$ determined from $m_\rho$.  These scales,
also listed in the table, have been previously reported by one of us
from a fit to the hadron spectrum \cite{gklat93}. We refer to these
below as the ``$m_\rho$ scales''.

We also give the values of $\alpha_\MSbar$ needed in the
one-loop matching between lattice and continuum operators.  These are
obtained using Eqs.~(\ref{eq:alpha1}) and (\ref{eq:alpha2}) from the
plaquette values listed in the table.  Since in the end we run our
results to $2\;$GeV, a consistency check on the applicability of
tadpole improved perturbation theory is that the values of
$\alpha_\MSbar(2 \GeV)$ agree for different $\beta$.  The table shows
that they are in reasonable agreement, although there is a small systematic
increase as $a$ decreases.  

This systematic dependence on $a$ suggests using a different set of
scales, defined so that $\alpha_\MSbar(2 \GeV)$ is independent of $a$.
We choose $\alpha_\MSbar(2 \GeV)=0.190$, and the resulting scales
are listed in the last row of the table.
We refer to these below as the ``$\alpha$ scales''.
We use these scales to determine our central values for matrix elements, 
and use the $m_\rho$
scales to estimate the error introduced by the uncertainty in determining $a$.

Our method for calculating $B$-parameters was developed for
calculating $B_K$ with staggered fermions \cite{bkprl}. It has also
been used for Wilson fermions \cite{bkwold}. A detailed description of
the latter application, and the extension to other operators is given
in Ref. \cite{bkwnew}. Here we confine ourselves to a brief
description of the features which are special to the present work.

The essential feature of the method is the use of wall sources
(i.e. sources confined to a single time-slice and extended over the
entire timeslice), which are designed to create particles with
specific quantum numbers.  In the current applications we want to
create only the lattice pseudo-Goldstone pion at rest, and this can be
accomplished using a linear combination of two wall sources, as
explained in Ref.~\cite{oldhm}.  The only other states created are rho
mesons and excited pions ($\pi'$).  Both of these are, however,
considerably more massive than the pseudo-Goldstone pion, and their
contribution can be largely removed by moving far enough away from the
source in Euclidean time.

The basic method is then to calculate ratios such as
\begin{equation}
B_K(t) =
{3 V_y \vev{W(t_{1})\sum_{\vec y}\CQ_K(\vec y,t) W(t_{2})} 
\over
8\vev{W(t_{1}) \sum_{\vec y'} A_4(\vec y',t)}
\vev{\sum_{\vec y''} A_4(\vec y'',t) W(t_{2})} }
\,.
\label{eq:bklatdef} 
\end{equation}
Here the operators are the
lattice versions of the continuum operators obtained after the
matching explained in the previous section has been carried out.
The wall sources are denoted by $W$ and are located at times $t_1$ and
$t_2$. The operators are placed at an intermediate time $t$
satisfying $t_1 \ll t \ll t_2$, and are summed over the $V_y$
spatial hypercubes.
Finally, the expectation values in Eq.~(\ref{eq:bklatdef}) are 
averages over quenched configurations.

The expression in Eq.~(\ref{eq:bklatdef}) is designed so that for large
enough $t-t_1$ and $t_2-t$ it is independent of $t$, and gives $B_K$
directly.  The point is that the exponential factors from Euclidean
time evolution cancel if only a single state, here the
pseudo-Goldstone pion, contributes.  This has to be true separately
for the numerator and denominator, i.e. both must exhibit a
``plateau'' over intermediate times in which they are independent of
$t$.  Thus the signal can be improved by averaging the numerator
and denominator over the time-slices in this plateau.  This is
what we do in practice, using the same range of times for both
numerator and denominator.

This method has several positive features.  First, it involves no
fitting.  Second, statistical errors are reduced by directly
calculating $B$-parameters rather than the matrix elements themselves.
Third, the ability to average over all the spatial points and over a
number of time slices improves the statistics.  This is possible
because we know the propagator from the wall sources to all points in
the lattice.  For the same reason, we can calculate matrix elements of
non-local operators without additional quark propagators.  And,
finally, we can use the same set of quark propagators to calculate
matrix elements with different choices of lattice operator.  The
latter two features are particularly important for staggered fermions
where most operators are non-local. If we placed the source of the
propagators at the position of the operators we would need $2^4$
sources for unsmeared operators ($4^4$ for smeared operators).

To calculate statistical errors, we use single elimination jackknife
in the following manner.  On each jackknife sample, 
we first match at the scale $q^*$, 
then linearly interpolate the results to physical kaon mass, 
and then evolve to the final scale $\mu=2\,$GeV,
and, finally, take the ratio which defines the $B$-parameter.  
The error is then obtained from the variation between samples.

We end this section with some details specific to our particular lattices.
\begin{itemize}
\item
At $\beta=6$, we use wall sources in Coulomb gauge,
while on the other lattices the sources are in Landau gauge.
The choice of gauge should have no impact on the final result
(as long as the non-locality introduced by the gauge fixing does not
extend from the source to the operator \cite{oldhm})
but might affect the statistical errors.
\item
At $\beta=6$, we use periodic boundary conditions (PBC) in space,
and Dirichlet bounday conditions in time.
We place the wall sources next to the boundary, i.e. at
$t_1=0$ and $t_2=39$.
The merits of this procedure have been discussed in Ref.~\cite{oldhm}.
We only note here that we use a plateau region of $t=10-29$.
\item
The lattices at $\beta=6.2$ and $6.4$ have too short an extent in time
to follow the method adopted at $\beta=6$.  Instead we use PBC in all
four directions, having first periodically doubled the lattice in
time.  We place wall sources at $t_1=0$ and $t_2=72$, and use the
plateau region $t=32-40$.  Note that the propagator with source at
$t_1=0$ ($t_2=72$) is, due to the PBC and doubling of the lattice, the
same as if the source was at $t_1=48$ ($t_2=24$). Because of this,
and the fact that each source produces pions propagating
in both forward and backward directions, we can make a second 
measurement with the plateau region at times $t=8-16$. 
We treat these two results as independent in our jackknife error analysis. 

To study possible sources of systematic errors due to 
contamination from excited states or ``off-shell'' matrix elements
we have also made the following measurements. 
Taking the same two wall sources to lie at $t_1=0$ and $t_2=24$
we use the region $t=32-40$ to obtain an estimate
of the ``off-shell'' matrix elements, i.e. those
in which both of the pions approach the operator from the same side. 
Placing the sources at $t_1=0$ and $t_2=48$,
and considering the plateau region at $t=23-24$,
we get an estimate of the on-shell matrix elements
having a greater contamination from excited states,
but a smaller contribution from off-shell matrix elements. 
Further discussion of these constructions and the associated 
systematic errors is given in Ref.~\cite{sslat90}.
In practice, we find that the various methods yield very similar results for
the $B$-parameters we consider, and that these sources of
error are considerably smaller than others as discussed later.\footnote{%
The same is not true for the auxiliary parameters $B_A$ and $B_V$
defined in Ref.~\cite{sschlogs}. For these we have no useful results
at $\beta=6.2$ and $6.4$.}
Thus we give no further details.
\end{itemize}

\section{Results for $B_K$}
\label{sec:bk}

Using the methods described in the previous two sections we extract 
the continuum $B_K$ at $\mu=2\GeV$ for each lattice spacing.
The only feature of the analysis not discussed above
is the interpolation to the physical kaon mass.
This we do by fitting $B_K$ itself (for our three mass points)
to a linear function of the squared lattice kaon mass, $m_{K,\LATT}^2$.
This is reasonable for staggered fermions because chiral
symmetry constrains $B_K$ to have the same form 
as in the continuum and in particular to be finite
in the chiral limit \cite{sstasi,sschlogs}.
For degenerate quarks, the explicit form is
\begin{equation}
B_K = B \left[ 1 - 3 y\ln y + b y + O(y^2) \right] \,,
\label{eq:bkchpt}
\end{equation}
where $y=m_K^2/(4\pi f_\pi)^2$ is the usual chiral expansion parameter. 
For the our range of $m_{K,\LATT}^2$ the $y\ln y$ contribution
is well represented by a linear function.
We note in passing that for Wilson fermions there is, in general, an
additional term proportional to $a/y$ in Eq.~(\ref{eq:bkchpt}) 
because of the explicit breaking of chiral symmetry.

\begin{figure}
\centerline{\hbox{\epsfysize=4.truein\epsfbox{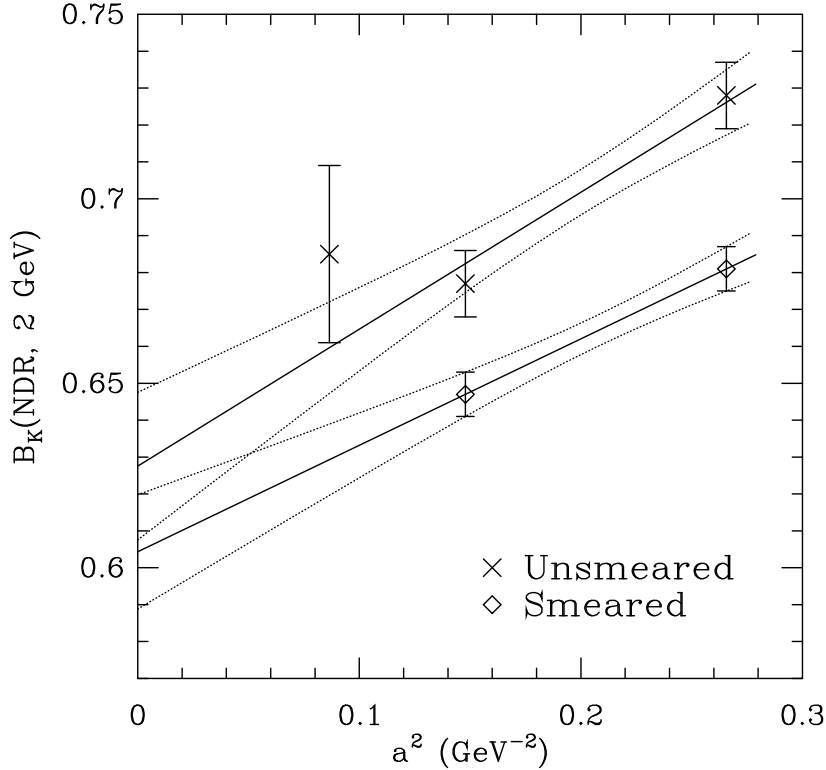}}}
\caption{Results for $B_K$ including extrapolation to the continuum limit.}
\label{fig:bkres2}
\end{figure}

We present results only for the $\alpha$ scales,
i.e. the lattice spacings determined by
requiring that $\alpha_\MSbar(2\GeV)=0.190$. 
These results are collected in
Table~\ref{tab:bkres2}, and displayed in Fig.~\ref{fig:bkres2}.
We include the results of an extrapolation to
the continuum limit assuming quadratic dependence on $a$.
Note that we do not have results for smeared operators at $\beta=6.4$.
Our results are not extensive
enough either to test whether the dependence on $a$ is indeed quadratic, 
or whether terms of higher order than quadratic are needed for our range
of lattice spacings.  The best confirmation of the validity of the
quadratic dependence comes from the work of the JLQCD collaboration,
who have more extensive results than ours for both the unsmeared operator
and its gauge-invariant version \cite{jlqcd96}.

\begin{table}
\renewcommand{\arraystretch}{1.2}
\begin{center}
\begin{tabular}{ccrrrr}
\hline
Operator & $q^*$	& $\beta=6.0$ 	& $\beta=6.2$	& $\beta=6.4$ & $a=0$\\
\hline

UN &$1/a$
		& 0.728(9)	& 0.677(9)	& 0.685(24) 	& 0.628(20)\\
UN &$\pi/a$
		& 0.725(9)	& 0.669(9)	& 0.677(24)	& 0.614(20)\\
\hline
SM &$1/a$
		& 0.681(6)	& 0.647(6)	&		& 0.604(15)\\
SM &$\pi/a$
		& 0.694(6)	& 0.655(6)	&		& 0.606(15)\\
\hline
Ratio UN/SM	&$1/a$	& 1.069(11)	& 1.046(13)	&	& 1.018(32)\\
Ratio UN/SM	&$\pi/a$& 1.044(11)	& 1.021(14)	&	& 0.991(34)\\
\hline
\end{tabular}
\caption{Results for $B_K(\NDR,2\GeV)$ at the physical kaon mass, 
using the $\alpha$ scales.
``UN'' and ``SM'' refer to unsmeared and smeared operators, respectively.}
\label{tab:bkres2}
\end{center}
\end{table}

We can use our results to estimate the systematic errors arising
from the choice of lattice spacings,
the truncation of perturbation theory in the matching factors,
and the contamination from excited states and off-shell matrix elements.

We begin with the dependence on the choice of lattice spacings.
This we estimate by comparing the results using the $\alpha$ scales
(listed in Table~\ref{tab:bkres2})
to those obtained using the scales determined from $m_\rho$.
The values of $B_K$ at each lattice spacing change by 
no more than $0.002$, and the
maximum change in the extrapolated value is $0.004$.
Thus we take $\pm 0.004$ for our estimate of this systematic error.
This is much smaller than the statistical errors.

The error due to the truncation of perturbation theory 
in the matching factors can be estimated in two ways.
The first, discussed in sec.~\ref{sec:theory}, 
uses the $q^*$ dependence of the result.
We take the error to be the difference between
the extrapolated results with $q^* a=1$ and $q^* a=\pi$.  
The table shows that this is comparable to the statistical errors.
The largest difference is that for the unsmeared operators,
and we take this for our estimate, yielding $\pm 0.014$.
This 3\% error is a reasonable estimate for a two-loop correction given
that the one-loop matching corrections for unsmeared and smeared operators
are $\sim 5-15\%$.
Our estimate of this error turns out to be the same as that quoted
in our preliminary result \cite{sslat93}, although the method we use here
is more reliable, as explained in sec.~\ref{sec:theory}.

The second way of estimating the perturbative error is to compare the results
using the two types of operator.  
If the matching factors were correct then they 
should yield the same result in the continuum limit.
We estimate the error as half the difference between the smeared and unsmeared
results at $q^* a=1$, and thus obtain $\pm 0.012$. 
In fact, this is likely to be an overestimate 
of the difference in the results from the two operators. 
This is because there is an extra data point at $\beta=6.4$ for
the unsmeared operators, and, as can be seen from Fig.~\ref{fig:bkres2},
this shifts the extrapolated result away from that of the smeared operator.
Perhaps a better way of estimating the error is to take the ratio of the
results using the two operators, and extrapolate this to the continuum limit.
This removes the data point at $\beta=6.4$
(since we have no smeared result there), 
and also accounts for the correlations between 
the results for the two operators.
The results for this ratio are given in Table~\ref{tab:bkres2}, 
and show that although the results from the two operators are
significantly different at finite $a$,
they are consistent in the continuum limit.
Because of this, we take our estimate of the perturbative
error from the dependence on $q^*$. 

Finally, we discuss the errors due to contamination in the
matrix elements from excited states and off-shell contributions.
The only excited state which is allowed to
contribute by the symmetries is the $\pi'$, i.e. the radially excited pion.  
There are no contributions from $\rho$ mesons 
(a point not realized in Ref.~\cite{sslat93}).  
As mentioned in Section~\ref{sec:details}, the size of the first effect can be
estimated by comparing the results with the two sets of sources
($t_1=0$, $t_2=72$ versus $t_1=0$, $t_2=48$),
while off-shell contamination can be estimated by studying $B_K(t)$ in 
regions where the off-shell contributions dominate.  
We find that the combined shift in $B_K$, 
after averaging the results from the two operators, 
and extrapolating to the continuum limit, is $\approx -0.003$.  
Since this estimate is approximate and small, 
we do not include this shift in our final result, 
but instead include it as part of the above overall systematic error.

Putting this all together, we can now quote our final result.
For the central value we use $q^*=1/a$,
and take the average of the results from the two operators.
We use the larger of the statistical errors (that for the unsmeared operators).
And we estimate the overall systematic error by combining linearly 
those from the choice of lattice scales ($0.004$), 
from the truncation of perturbation theory ($0.014$), 
and from the contaminations ($0.003$).
Thus we quote
\begin{equation}
B_K(\NDR,2\GeV) = 0.62 \pm 0.02 ({\rm stat}) \pm 0.02 ({\rm syst}) \,.
\end{equation}
This result is consistent with our preliminary number quoted in 
Ref.~\cite{sslat93} ($0.616\pm 0.020 \pm 0.017$), although
the precise agreement is somewhat fortuitous given that our method
of matching has been improved.

Our results for the unsmeared operator can be checked by comparing
them to those from the JLQCD collaboration \cite{jlqcd96}.  They have
results at $\beta=6.0$, $6.2$, and $6.4$, on lattices of the same
spatial sizes as ours, but with statistical errors two or three times
smaller.  They also have results on larger lattices at $\beta=6$ and
$6.4$, and at smaller values of $\beta$.  A direct comparison is
possible because they use a method of matching and determining
$\alpha_\MSbar$ which is very close to ours if we set $q^*=1/a$.  
At $\beta=6$ and $6.2$, our results are larger by about $0.025$, a two
standard deviation difference.  At $\beta=6.4$ our number is
consistent with theirs from a $32^3$ spatial lattice.  They find,
however, that $B_K$ decreases on larger lattices, suggesting that our
number at $\beta=6.4$ may be afflicted by finite size errors.
Nevertheless, our extrapolated value is only $1.5$ standard deviations
(i.e. $0.03$) above theirs.
This comparison gives us confidence that our results are correct within
the quoted errors, and in particular 
that our procedure of doubling the lattice in
the time direction has not introduced additional systematic errors.

The JLQCD collaboration has also used the gauge-invariant
version of the unsmeared operator.  Averaging this with
the Landau gauge operator, they quote a preliminary result
$B_K(\NDR, 2\GeV) = 0.587 \pm 0.007 ({\rm stat}) \pm 0.017 ({\rm syst})$,
or adding errors in quadrature: $B_K = 0.59 \pm .02$.
The agreement with our result of $B_K = 0.62 \pm .03$ is
gratifying, since the operators we use have entirely different
perturbative and power corrections.\footnote{%
We can improve the agreement by dropping
our unsmeared data point at $\beta=6.4$, which, as noted above,
might be afflicted with finite size errors.}

We note that the both calculations are systematics limited, and
that of the systematic errors, the most important one quoted by
JLQCD is estimated by comparing results for different operators.  
In their preliminary report, 
JLQCD ascribes such differences to terms of order
$\alpha_\MSbar(\mu)^2$, where $\mu = 2\GeV$ is
the conventional scale at which the answer is quoted.  
On this point we disagree in principle.  We certainly agree that
errors of order $\alpha_\MSbar(\mu)^n$
are introduced when one uses $n$-loop evolution to the final
scale, but these corrections are universal and should not
appear as differences between lattice operators.
If, for example, we compare our one-loop corrected smeared and
unsmeared operators, the connection between the two is
\begin{equation}
\CQ_K^\UN(\mu) = \CQ_K^\SM(\mu) \left\{
1 + O[\alpha_\MSbar(q^*)^2] + O(a^2) \right\}
\,,\label{eq:matchUNtoSM}
\end{equation}
where we have taken the same $q^*=K/a$ in renormalizing
both operators.\footnote{%
Using a different choice of $q^*$ for the two operators would lead to
an additional factor coming from the evolution between the two scales.
However, since this factor tends to unity in the continuum limit, our conclusion 
is unaltered.}
Since $\alpha_\MSbar(K/a)^2$ vanishes as $a\rightarrow0$, 
we conclude that when correctly extrapolated, 
the operators should give the same result at $a=0$.
Of course one would need rather precise data to make a
fit including the $\alpha_\MSbar(K/a)^2$ term,
but in principle it could be done if more precision were required.  
If one uses a simple $a^2$ extrapolation, however, the $\alpha^2$ term
in Eq.~(\ref{eq:matchUNtoSM}) will not extrapolate to zero, but instead
to an artifact of size $\alpha_\MSbar(q^*)^2$.

\section{Results for $B_7^{3/2}$ and $B_8^{3/2}$}
\label{sec:b78}

We have evaluated $B_7^{3/2}$ and $B_8^{3/2}$ using almost the 
same method as for $B_K$.
The only difference concerns the extrapolation to the physical kaon mass. 
This we have done separately for matrix elements appearing
in the numerator and denominator of the definitions Eqs.~(\ref{eq:B7def})
and (\ref{eq:B8def}), prior to the evolution from $q^*$ to $\mu=2\GeV$.

Our results are summarized in Table~\ref{tab:b78res}.  The most
striking feature is the very strong $q^*$ dependence of the results for
unsmeared operators.  Indeed, as one goes from $q^*=\pi/a$ to
$1/a$, which causes $\alpha_\MSbar(q^*)$ to increase by roughly $40\%$, 
$B_{7,8}^{3/2}$ change sign because the negative one-loop matching
contribution exceeds the tree-level contribution. 
Clearly, we cannot use one-loop matching for the unsmeared operators.
We stress that the large perturbative corrections 
for unsmeared operators do not invalidate the results for smeared operators. 
There will always be choices of discretization procedure
for which the perturbative series is poorly convergent at the lattice
spacing one is working.

For completeness we mention that we expect a similar problem to render
the gauge-invariant unsmeared operators unsuitable for a calculation
of $B_{7,8}^{3/2}$. The dominant contribution to the matrix elements
comes from the PP part of the operator.
Since this part of the operator is local, and does not require gauge
links, it is the same as the corresponding part of the unsmeared operator.
But it is this part of the unsmeared operator which leads to the bulk of
the large one-loop matching corrections seen in sec.~\ref{sec:theory}.
Although in principle it is possible these large corrections
will be canceled by the as yet uncalculated contributions
from the other parts of the gauge-invariant operator, 
this seems unlikely in practice.

\begin{table}     
\renewcommand{\arraystretch}{1.2}
\begin{center}
\begin{tabular}{lcrrr}
\hline
Operator	& $q^*$		& $\beta=6.0$ 	& $\beta=6.2$	& $a=0$\\
\hline\\[-.5em]
\multicolumn{5}{c}{$B_7^{3/2}$} \\
\hline
Unsmeared 	& $1/a$ 	&-1.951(09)	&-1.097(15) \\
Unsmeared 	& $\pi/a$	& 0.606(06)	& 0.645(06) \\
\hline
Smeared 	& $1/a$		& 0.989(05)	& 0.823(16)	& 0.615(30)\\
Smeared 	& $\pi/a$	& 1.085(06)	& 0.903(14)	& 0.674(32)\\
\hline\\[-.5em]
\multicolumn{5}{c}{$B_8^{3/2}$}\\
\hline
Unsmeared 	& $1/a$ 	&-1.486(11)	&-0.689(15) \\
Unsmeared 	& $\pi/a$	& 0.822(05)	& 0.864(05) \\
\hline
Smeared 	& $1/a$		& 1.240(06)	& 1.030(16)	& 0.766(37)\\
Smeared 	& $\pi/a$	& 1.288(06)	& 1.076(17)	& 0.810(39)\\
\hline
\end{tabular}
\caption{Results for $B_7^{3/2}(\NDR,2\GeV)$ and $B_8^{3/2}(\NDR,2\GeV)$,
at the physical kaon mass, using $\alpha$ scales.}
\label{tab:b78res}
\end{center}
\end{table}

The situation is much improved for the smeared operators---the $q^*$
dependence, while more significant than for $B_K$, is only at the
10\% level.  This is a much larger uncertainty than the statistical
errors or that due to the choice of lattice spacings, or that from the
contamination by excited states or off-shell matrix elements.  Thus we
do not give details concerning these other uncertainties.  For our
final results, we quote
\begin{eqnarray}
B_7^{3/2}(\NDR,2\GeV) &=& 0.62 \pm 0.03 {\rm (stat)} \pm 0.06 {\rm (syst)}\,,\\
B_8^{3/2}(\NDR,2\GeV) &=& 0.77 \pm 0.04 {\rm (stat)} \pm 0.04 {\rm (syst)}\,,
\end{eqnarray}
where the systematic error is our estimate
(based on the $q^*$ variation)
of the uncertainty due to the truncation of perturbation theory.

It is interesting to compare our staggered results with
those from Wilson fermions.
While a continuum extrapolation is not yet available, 
the results at $\beta=6.0$ \cite{bkwnew} are 
\begin{eqnarray}
B_7^{3/2}(\NDR,2\GeV,{\rm Wilson}) &=& 0.58 \pm 0.02 \pm 0.07 \,,\\
B_8^{3/2}(\NDR,2\GeV,{\rm Wilson}) &=& 0.81 \pm 0.03 \pm 0.03 \,.
\end{eqnarray}
The central values and error bars have been determined
in the same way as in this paper.
Obviously the agreement is already rather good, and indicates
indirectly that the $O(a)$ errors in the Wilson case 
are not particularly large.
This is in contrast to the case of $B_K$, where the explicit
breaking of chiral symmetry disrupts the delicate cancellation
between the VV and AA matrix elements, and gives dramatically
large $O(a)$ errors. For $B_7$ and $B_8$
the result is dominated by the PP matrix element, with no
particular constraint from chiral symmetry, presumably yielding
more moderate $O(a)$ errors.

\section{Non-perturbative results for matching constants}
\label{sec:zp}

It is clear from the results of the previous two sections that
using finite order perturbation theory to match lattice and
continuum operators is an important source of uncertainty in results
for matrix elements. In this section we investigate the accuracy
of perturbative matching factors by calculating 
some of them non-perturbatively. In particular, we are able to
assess the accuracy of the perturbative matching factor for the
quark mass, $Z_m$, which is an important ingredient in determining
continuum quark masses.

The basic idea is simple, and has been applied extensively with Wilson
fermions. A given continuum operator can be discretized in different
ways, each discretization having an associated matching factor.
Only if these matching factors are chosen correctly will the different
choices yield the same matrix elements. This allows a non-perturbative
determination of the ratio of matching factors. Note that in such
ratios the anomalous dimension factors cancel, implying that the ratios are finite
functions of the lattice coupling.

In this section, we apply this idea to the pseudoscalar and axial densities.
It could, in principle, be applied also to four-fermion operators such
as $\CQ_K$, but this requires studying a matrix mixing problem, and thus
using several external states, and is beyond the scope of the present work.
We only consider operators for which the matching is diagonal.

Our notation in this section differs from that used above.
We use $P^\UN$, for example, to refer to the bare lattice operator
constructed of unsmeared fields,
\begin{equation}
P^\UN = (1/\sqrt{N_f})
\bar\chi_S^\UN \overline{(\gamma_5\otimes\xi_5)} \chi_D^\UN
\,.
\end{equation}
In other words, we do not include the matching factor in the definition
of $P^\UN$. The matching equation becomes [cf. Eq.~(\ref{eq:contlatt})]
\begin{eqnarray}
P^\CONT &=& Z_P^\UN P^\UN [1 + O(a^2)] \\
Z_P^\UN &=& 1 + 
{\alpha_\MSbar(q^*)\over 4\pi} \big(-\gamma_P^{(0)} \ln(q^*a) + c_P^\UN \big)
+ O(\alpha^2) 
\,.
\end{eqnarray}
Here we have used the fact that for the matrix elements of
interest the corrections are quadratic in the lattice spacing.
Similar definitions apply to $P^\SM$, $A_\mu^\UN$ and $A_\mu^\SM$.
We also need to introduce the gauge-invariant version of the axial
current $A_\mu^{\rm GI}$, which is the unsmeared current with appropriate
gauge links included. Note that $P^{\rm GI} = P^\UN$, i.e. the unsmeared
pseudoscalar density is already gauge invariant since it is local.

The quantities we determine non-perturbatively are
$Z_P^\SM/Z_P^\UN$, $Z_A^\SM/Z_A^\UN$, and $Z_A^\UN$.
The first we obtain starting from the result
\begin{equation}
\vev{0|P^\CONT |\bar{K}^{0}} = 
\vev{0| Z_P^\UN P^\UN |\bar{K}^{0}_{G}} [1 + O(a^2)] =
\vev{0| Z_P^\SM P^\SM |\bar{K}^{0}_{G}} [1 + O(a^2)] 
\,
\end{equation}
from which we find the non-perturbative estimate
\begin{equation}
{Z_P^\SM \over Z_P^\UN} = 
{ \vev{0| P^\UN |\bar{K}^{0}_{G}} \over   \vev{0| P^\SM |\bar{K}^{0}_{G}}}
\left[1 + O(a^2) \right] 
\,.
\end{equation}
The equations for $Z_A^\SM/Z_A^\UN$ are identical
except that $P$ is replaced by $A_4$.
The one-loop perturbative results for the relevant ratios are 
\begin{eqnarray}
{Z_P^\SM \over Z_P^\UN} &=&
\left[ 1 + {\alpha_\MSbar(q^*)\over 4\pi} (c_P^\SM-c_P^\UN) \right] 
= 1 + {\alpha_\MSbar(q^*)\over 4\pi} \times 30.2532   \,, \\
{Z_{A}^\SM \over Z_A^\UN} &=&
\left[ 1 + {\alpha_\MSbar(q^*)\over 4\pi} (c_A^\SM-c_A^\UN) \right] 
= 1 - {\alpha_\MSbar(q^*)\over 4\pi} \times 1.9384   \,. 
\end{eqnarray}

The determination of $Z_A^\UN$ proceeds slightly differently.
We note that the gauge-invariant axial current is partially conserved
on the lattice, implying that $Z_A^\GI=1$. Thus we could determine $Z_A^\UN$ 
by taking ratios of matrix elements of $A_\mu^\GI$ to those of $A_\mu^\UN$. 
While we have not calculated matrix elements using $A_\mu^\GI$,
we do have available the matrix elements of its divergence,
$\partial_\mu A_\mu^\GI =2 m P^\GI = 2 m P^\UN$.
This is sufficient as long as the matrix element involves
non-zero momentum transfer.
Note that the lattice partial conservation equation
is exact as long as we use the appropriate lattice derivative \cite{toolkit}.
Putting this all together we arrive at
\begin{equation}
{1 \over Z_A^\UN} = \left(
{- \sinh(m_K)  \vev{0| A_4^\UN |\bar{K}^{0}_{G}} \over
2 (m_q/u_0)  \vev{0| P^\UN |\bar{K}^{0}_{G}}} \right)
\left[ 1 + O(a^2) \right]
\,,
\label{eq:nonpertZA}
\end{equation}
where $m_q/u_0$ is tadpole improved quark mass, and $m_K$ is the
mass of the $K^{0}_{G}$. All quantities on the r.h.s. of this equation
are in lattice units.
We have used $\sinh(m_K)$ on the r.h.s., rather than $m_K$ itself,
because if we replace $A_4^\UN$ by $A_4^\GI$ then the ratio on the
r.h.s. is exactly equal to unity.
In other words, $\sinh(m_K)$ is the appropriate kinematical factor for the
exactly conserved current.
We choose to keep it for the unsmeared current in the hope that it will
reduce the size of the $O(a^2)$ terms.
The perturbative expression to which Eq.~(\ref{eq:nonpertZA}) should be
compared is
\begin{equation}
{1 \over Z_A^\UN} = 
\left[ 1 + {\alpha_\MSbar(q^*)\over 4\pi} {4 \over 3} (\pi^2-9.17479) \right]
\,.
\end{equation}
Note that if we had used the average link in Landau gauge to determine
$u_0$, rather than the average plaquette, 
then $Z_A^\UN=1$ at one-loop order.

We determine the required ratios of matrix elements using 
the quantities previously used to determine the vacuum saturation
approximants appearing in the $B$-parameters. 
For example, consider the ratio
\begin{equation}
R_P(t) =
{\vev{W(t_{1}) \sum_{\vec y} P^{\UN}(\vec y,t)}
\vev{\sum_{\vec y'} P^{\UN}(\vec y',t) W(t_{2})} 
\over
\vev{W(t_{1}) \sum_{\vec y} P^{\SM}(\vec y,t)}
\vev{\sum_{\vec y'} P^{\SM}(\vec y',t) W(t_{2})} }
\,.
\label{eq:ratdef} 
\end{equation}
For $t_1 \ll t \ll t_2$ this should be independent of $t$ and
gives directly
$(Z_P^\SM/Z_P^\UN)^2$ aside from $O(a^2)$ corrections.
We average $R_P(t)$ over the same plateau regions as for the $B$-parameters.
Similar ratios are used for the other quantities.

The resulting data for $Z_P^\SM/Z_P^\UN$ and $Z_A^\SM/Z_A^\UN$ are
well represented by a linear function of $m_K^2$.
This dependence on $m_K^2$ is an $O(p^2a^2)$ discretization error,
because any physical dependence cancels in the ratio of matrix elements,
We remove this error by extrapolating to the chiral limit.
We do a similar extrapolation for $1/Z_A^\UN$, 
although the dependence on $m_K^2$ is much weaker,
presumably because of the $\sinh(m_K)$ factor in Eq.~(\ref{eq:nonpertZA}).

Our non-perturbative 
results for the ratio of smeared to unsmeared matching factors after
chiral extrapolation are collected in Table~\ref{tab:ratres}.
What is most striking is the substantial dependence on lattice spacing,
particularly for the ratio of $Z_P$'s.  This is due to a combination
of $O(a^2)$ discretization errors and the variation of the
perturbative matching factors which depend on $g^2(a)$.
To analyze these results we assume the following form for the ratios
\begin{equation}
{\rm RATIO}({\rm non\!-\!pert}) =
{\rm RATIO}({\rm one\!-\!loop};q^*) + a^2 \Lambda^2
\,,
\label{eq:ratioform}
\end{equation}
where the one-loop results are given above, 
and $\Lambda$ is an unknown constant.
In other words we ignore completely higher powers of $a$,
and assume that higher powers of $\alpha$ are well represented by the
appropriate choice of $q^*=K/a$.
The difference RATIO({non-pert})$-$RATIO({one-loop})
should then, for the right choice of $q^*$, extrapolate to zero in the
continuum limit.  
Conversely, one could regard this procedure as providing an approximate
non-perturbative definition of $q^*$.
To show the individual variations in the perturbative\footnote{%
Note that we are here making use of the fact that our extrapolation to the
continuum limit does not remove terms which vary logarithmically.
This is an example of the problem discussed in sec.~\ref{sec:bk}.}
and non-perturbative results we give, in Table~\ref{tab:ratres}, 
the continuum value for each obtained by linear extrapolation in $a^2$.  
We do this for our two standard choices $q^*a=1$ and $\pi$.
What we find remarkable is that, modulo the simplifying
assumption of Eq.~(\ref{eq:ratioform}), 
the non-perturbative and perturbative predictions agree if
we use $q^* \approx 1/a$ but not for $q^*\approx \pi/a$.  The
discrepancy for $q^*=\pi/a$ is particularly significant for
$Z_P^\SM/Z_P^\UN$. 


\begin{table}
\renewcommand{\arraystretch}{1.2}
\begin{center}
\begin{tabular}{lllll}
\hline
Quantity	  & Method      	&$\beta=6.0$ 	& $\beta=6.2$	& $a=0$ \\
\hline
		  & Non-pert    	& 1.131(5)	& 1.060(1)	& 0.972(6) \\
$Z_A^\SM/Z_A^\UN$ & Pert$(q^*=1/a)$ 	& 0.970         & 0.973         & 0.977	\\
		  & Pert$(q^*=\pi/a)$	& 0.979		& 0.981		& 0.983 \\
\hline
		  & Non-pert    	& 2.245(11)	& 1.861(6)	& 1.380(19) \\
$Z_P^\SM/Z_P^\UN$ & Pert$(q^*=1/a)$     & 1.462         & 1.416         & 1.359	\\
		  & Pert$(q^*=\pi/a)$	& 1.323		& 1.301 	& 1.272 \\
\hline
\end{tabular}
\caption{Non-perturbative and perturbative results for ratios of
matching constants.  Quadratic extrapolations to 
$a=0$ use $\alpha$ scales.
}
\label{tab:ratres}
\end{center}
\end{table}

The results for $1/Z_A^\UN$, given in Table~\ref{tab:ZAres},
behave very differently. There is very little dependence on the
lattice spacing---presumably because the unsmeared current is
very similar to the gauge-invariant current for which all the
numbers in the table would be unity independent of lattice spacing.
In fact, the errors are such that an extrapolation to $a=0$ is
not useful, and so we compare our results to perturbation theory
at each lattice spacing. The results are reasonably
consistent, but we cannot distinguish between different values of
$q^*$ in this case.

\begin{table} 
\renewcommand{\arraystretch}{1.2}
\begin{center}
\begin{tabular}{llll}
\hline
Method		& $\beta=6.0$ 	& $\beta=6.2$	& $\beta=6.4$ \\
\hline
Non-perturbative&   0.978(09)	& 1.010(09)	& 0.998(20) 	\\
Pert ($q^*=1/a$)&   0.986	& 0.987		& 0.988		\\
Pert ($q^*=\pi/a$)& 0.990	& 0.991		& 0.991		\\
\hline
\end{tabular}
\caption{Results for $1/Z_A^\UN$.}
\label{tab:ZAres}
\end{center}
\end{table}

An important application of the above results is to estimate
the reliability of the one-loop result for the matching factor $Z_m$.
This factor converts the lattice results for quark
masses to a continuum scheme like $\MSbar$, 
as discussed in Ref. \cite{GB}.  The perturbative result, after tadpole 
improvement is
\begin{equation}
Z_m(\mu=q^*) = {1 \over Z_P^\UN(\mu=q^*)} = 
1 - 
{\alpha_\MSbar(q^*)\over 4\pi} (-\gamma_P^{(0)} \ln(q^*a) + c_P^\UN)
\,,
\end{equation}
where for simplicity we have chosen to consider the case where
the final scale $\mu$ equals the matching scale $q^*$.
The one-loop correction to $Z_m$ is large at typical lattice spacings.
For example, at $\beta=6$, and taking $q^*=1/a$, $Z_m=1.598$.
This suggests that higher order corrections may be important.
We can, however, rewrite $Z_m$ as
\begin{equation}
Z_m = \left({Z_P^\SM\over Z_P^\UN}\right)\left( {1 \over Z_P^\SM}\right)
\,.
\end{equation}
The results of Table~\ref{tab:ratres} show that the bulk of the perturbative
correction lies in the first factor, $Z_P^\SM/ Z_P^\UN$.
At $\beta=6$ it is $1.462$,
while the second factor is $1.136$ (again for $\mu=q^*=1/a$).
Thus, one would expect that the dominant source of higher order terms in $Z_m$
is the first factor, and that the uncertainty which they introduce could be
substantially reduced by obtaining a non-perturbative estimate of this
factor. We have attempted such an estimate above, 
with the preliminary conclusion that perturbation theory with 
$q^*=1/a$ works to within a few percent, aside from discretization errors. 
If we accept this result then we obtain the partly non-perturbative
estimate\footnote{%
It is not advantageous to directly use the non-perturbative results,
e.g. $Z_P^\SM/ Z_P^\UN=2.245$ at $\beta=6$. Doing so introduces additional
$O(a^2)$ errors which one would then have to remove by extrapolation.}
$Z_m = 1.462 \times 1.136 = 1.66$. The point we wish to
stress is that this is not very different from the one-loop estimate
of $1.60$.  In particular, the difference is much smaller than the
naive estimate of the two-loop contribution, $0.6^2=0.36$,
based on the assumption of geometric growth.
This analysis thus suggests that the
one-loop perturbative value of $Z_m$ used in the analysis of quark
masses is good to about $5\%$ for $\beta \ge 6.0$ provided one uses
$\alpha_\MSbar(q^* \approx 1/a)$.

\section{Conclusions}
\label{sec:conc}

In this paper we have presented a variety of results for weak matrix 
elements using staggered fermions. Our major focus has been on the
importance of using a variety of discretizations of
continuum operators. There are two reasons for doing so.
First, comparing results with different lattice operators gives 
an estimate of the uncertainty in the matching factors between
continuum and lattice operators. For $B_K$ this may
be the dominant source of error in future calculations,
aside from that due to quenching.
Second, for some operators the perturbative matching factors are not
convergent at present couplings, 
and so one must use different discretizations.
It turns out that the smeared operators have uniformly moderate
perturbative corrections. Using them we are able to obtain the
first results for $B_{7}^{3/2}$ and $B_{8}^{3/2}$
using staggered fermions.

Our results for the $B$-parameters confirm and extend existing lattice
results. In particular, for $B_K$ we
find that smeared operators give results consistent 
with those from unsmeared and gauge-invariant operators.
We confirm the low value found in our preliminary study \cite{sslat93},
a result which has been improved and extended by
the JLQCD collaboration \cite{jlqcd96}.
For $B_{7}^{3/2}$ and $B_{8}^{3/2}$
we find results consistent with those using Wilson
fermions. All these numbers are important inputs into analyses attempting
to constrain the CKM matrix.
It is encouraging that the errors we are considering are at the few percent
level. It is important to stress, however, that we are still using
the quenched approximation, and also working with a kaon composed
of degenerate quarks. For a discussion of the importance of these
approximations see Refs. \cite{sslat97,sstasi}.

As an offshoot of our study, we have calculated several ratios of 
matching factors non-perturbatively. These ratios are finite functions
of the lattice coupling, and thus allow a test of tadpole improved
perturbation theory. We find that one-loop perturbation theory works
well if we set the scale in the one-loop coupling,
$\alpha_\MSbar(q^*)$, to be $q^*=1/a$, but not for $q^*=\pi/a$.
This is consistent with the expectations of Ref.~\cite{lepage}.
This conclusion is, however, preliminary because we have results only
at two lattice spacings.
To convincingly disentangle discretization errors from perturbative corrections
will require precise results at several lattice spacings. 

We have used the results for ratios of
matching factors to make a partly non-perturbative
estimate of the size of the matching factor for the quark mass, $Z_m$. 
Our result is $\sim 5\%$ higher than the one-loop perturbative result. 
This is a small enough change that
it does not alter the essential conclusion of Ref. \cite{GB}, namely that
light quark masses are considerably smaller than previously thought.

Finally, we note that our results show many examples of significant
discretization errors. To make progress with simulations of full QCD,
where one is restricted to larger lattice spacings, it may be necessary
to improve the staggered fermion action.

\section*{Acknowledgements}

We gratefully acknowledge the support of DOE under the 
Grand Challenges allocation of computer time at the National Energy 
Research Supercomputer Center (NERSC).

\end{document}